# Optical-Wireless Hybridization in the Core-Access in 5G

Sudhir K. Routray, Habib H. Mohammed
Department of Electrical and Computer Engineering
College of Electrical and Mechanical Engineering
Addis Ababa Science and Technology University, Addis Ababa

*Abstract*—The fifth generation (5G) cellular communication system will have several advanced features, majority of which will come into existence for the first time. According to the ITU recommendations, 5G will provide minimum 3 GB/s end to end data rates under the static conditions and in the mobile condition the minimum data rate is 100 Mb/s. The latency will be brought down to 1 ms and the device densities will increase by many folds. These features cannot be provided in the pure wireless domain. Certainly, the support of the optical networks is essential to provide these advanced features in 5G communication. These features need special architectures. In this article, we provide advanced optical wireless hybrid architectures for 5G networks. We explain the importance of this architecture and also show the nodal configuration of these hybrid networks.

## I. INTRODUCTION

The features of proposed 5G cellular systems as set by ITU for are astronomical. It is intended to provide 10 Gb/s data rate to the end costumers in the static environment (minimum: 3 Gb/s, and maximum: 20 Gb/s). For the mobile environment, the minimum data rate is set at 100 Mb/s. In addition to the above, several other high performances are expected to be achieved in the 5G environment. These performances are not possible without a highly efficient optical core network. In addition to the core, optical fibers are to be present in the regional, metro and access areas. Preferably, in the majority of cases we need it up to the base stations. From the base stations to the end customers which is known as the last mile will be covered in the wireless domain. Therefore, appropriate optical wireless communication architectures are need for 5G.

5G communication systems is the most recently proposed cellular system that is expected to provide fiber-like experience in the wireless environment. The average data rates expected in the 5G environment is of the order of 10 Gb/s [1]. This data rate is currently available in the access area only through the optical access networks, also known as the passive optical networks (PONs). Making it available in the wireless form is a challenge in itself. With this huge end to end data rates, the trunk speeds have to be really gigantic (of the order of several Tb/s). This speed is only possible though the optical fibers. Therefore, 5G definitely needs the fiber support for its successful deployment. The data rates and spectral efficiency proposed for 5G are large enough to use the existing 3G and 4G spectrum. Therefore, new microwave spectrum has been suggested to be the right choice. The basic structure of 5G will be dependent on small cell access networks. For that, mm-waves (this is a part of microwave spectrum whose wavelengths are in the order of millimetres) have been proposed as the small cell data carriers. As the mm-waves are fast fading in nature, only the base station to the end customer communication would remain wireless. From the base station to the rest part of the core network the medium would be wired, preferably optical fibers. In [2], optical transport has been proposed as the main solution for the 5G deployment. The authors in this article proposed a fiber-based core and metro framework for the 5G and summarized the key defining factors for meeting the 5G requirements. 5G will be face the challenge to cater with an extremely diverse environment due its versatility and the high demand for the data rates. At the same time the network has to be very flexible, scalable and cost-effective which can enable service differentiation [2]. This work has outlined the WDM based solutions for 5G.

A combination of fiber and radio links for 5G has been proposed for small cells and moving cells in [3]. In such hybrid networks, overall resource management, latency control and power consumption related issues become very complex. In [3], the authors have proposed an overview of such complex scenario, and solutions for 5G mobile fronthaul using analog transmission of bidirectional seamless systems. They presented the challenges encountered in the heterogeneous small cell and moving cell networks. A proof of the associated concepts and demonstration is presented to support their proposed solution. They used long optical links to facilitate practical implementation of such hybrid networks for 5G. Microwave photonics is a leading technology for the practical deployment of 5G in the coming years [4], because it integrates both microwave and photonics. A perfect combination of these two technologies is required for the success of 5G. Microwave in the mm-wavelength range has been proposed for the last hop wireless communication in 5G, and optical communication is the only way the huge 5G trunk capacity can be provided. The authors provided a detailed review of the modern cellular radio access networks (RANs) and identified the segments where microwave photonics has significant roles. They also provided the advantages of integrated microwave photonics for cellular communications and proposed a fronthaul for the RAN for upcoming cellular networks. In [5], operators prospective are presented from the next generation cellular communication points of views. The authors emphasized on the optical communication based local access networks for the future radio access. Their idea on the optical access for the future radio access network is based on TDM-PON based small cell networks. Photonics technologies, especially microwave photonics has a significant role for the development of 5G [6]. It not only provides compact semiconductor devices in the microwave environment but also has the ability to provide high performance operational characteristics. In [6], all the proposed disruptive technologies for 5G are discussed. The authors presented in detail, how the microwave and photonic integration will affect the small cell architecture, mm-wave characteristics and the massive MIMO at a large scale. Fronthaul technologies have several ramifications and requirements. It is explained from the optical perspectives points of views in [7]. In [8], the authors presented an integrated architecture for both fronthaul and backhaul and they named it as xhaul. In this initiative they try to provide a common channel for the uplink and downlink at most part of the network except for the last mile.

In [9], main characteristics of mm-waves are presented for both short and medium range communications. Several practical measurements are carried out in both complex urban like terrain and simple planner terrain with fewer complexities [9]. In either of the cases, signal attenuation, multipath fading and other wave propagation characteristics are found to be suitable for high data rate communications. Therefore, the authors proposed mm-waves as the prime candidate for 5G access area spectrum. Small cells are essential for 5G as the fading characteristics for urban environments are really fast for mm-waves though they can be used for larger coverage for rural areas with planar terrain. In [10], the importance of small cell is justified. This is also validated using experimental findings and the performance dynamics of small cells. In [11], new 5G initiatives, the long term plans and the recent developments are presented for the associated research communities. It provides all the European activities carried under the 5G-PPP umbrella. 5G will be a very advanced access area technology. 5G Ran will have several differences from its legacy systems. In [12], main features of 5G RAN its design guidelines have been presented. In [13], the recent changing trends of optical communications have been presented. It shows that the optical technologies and wireless technologies follow almost the similar paths in their advancement. For instance all the advanced wireless techniques such as orthogonal frequency division multiplexing and space division multiplexing are being used in optical communications. Coherent communication is also very popular for long distance optical communications which is always very popular in wireless. Different types of architectural evolution are presented in [14]. The chronological evolutions of the mobile communications and their architectural changes are highlighted in this article. It shows the gradual changes in the architectures of cellular networks. Software defined networking is an essential feature of 5G which will segregate the services using different types of slices. In [15], software defined networking for 5G has been discussed considering a hybrid optical wireless framework. In [16], a software defined networking framework for optical networks has been presented. It provides the basic concepts and requirements for software defined optical networks. It is shown that the network costs can be reduced using software defined network management.

In this article, we present hybrid core and access networks for 5G which are hybrid optical wireless networks. We explain the justification for each of the cases presented. We provide different architectures for rural and urban areas as the demands are different.

The reminder of the article is organized in three sections. In Section II, we present the 5G cellular communication system architecture as proposed by ITU. In Section III, we present the core and access hybridization using both optical and wireless components. In Section IV, we conclude the paper with the main summary.

## II. 5G CELLULAR COMMUNICATION SYSTEMS

5G is envisioned as an advanced cellular communication system with several high performance characteristics. Depending on their domain of expertise, people explain and define 5G is different words. While the data rates and throughputs are the main parameters which are highlighted most of the time, other parameters too are important. Here we show a comparative figure with respect to 4G which can be easier to understand the 5G. The traffic in a typical 5G network will be around 10000 times the traffic of a similar sized 4G network. The number of devices in 5G will increase

TABLE I

MAIN ITU ROADMAP FEATURES OF 5G AND ITS MAIN PROPOSED TECHNOLOGIES FOR DEPLOYMENT

| # | Features | Range | Proposed Technologies |
|---|---|---|---|
| 1 | Data rate | 0.1 – 20 Gbps | Massive MIMO, MM wave spectrum |
| 2 | Spectral Efficiency | 4.5 | Advanced modulation, Full duplex, Massive MIMO |
| 3 | Data Processing | 10 Mb/s/m$^2$ | RAN vitalization, Small cells |
| 4 | Device Density | 1 million/km$^2$ | D2D, Small cells |
| 5 | Mobility | Up to 500 km/h | Heterogeneous network architectures |
| 6 | Transmission Delay | 1ms | D2D, Content caching close to users |
| 7 | Energy Consumption | 1 µJ per 100 bits | Massive MIMO, Core optical communications |

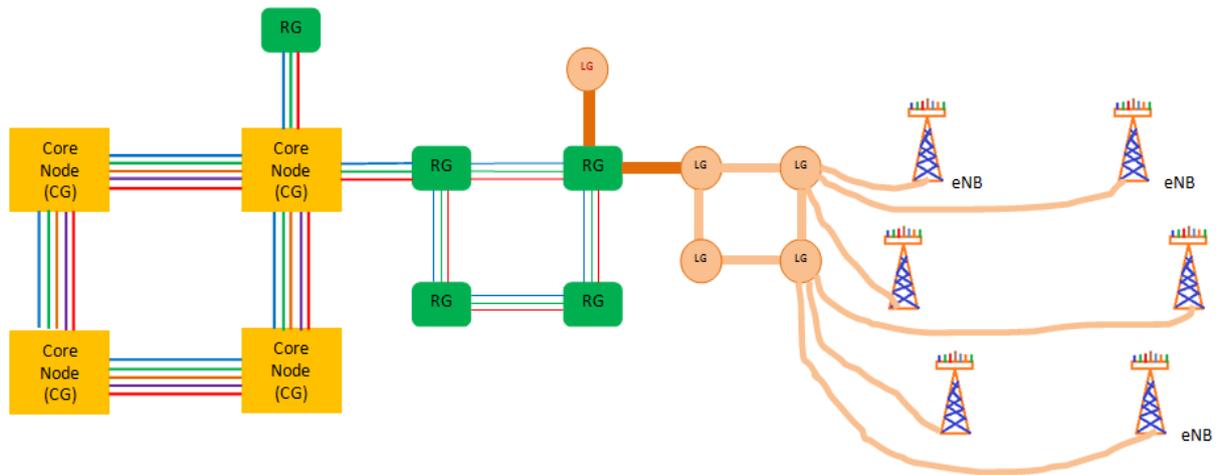
Fig. 1. 5G architecture for urban areas based on optical wireless hybrid communication systems. In this case, the whole architecture is optical except for the last mile.

in between 10 and 100 folds with respect to the corresponding 4G network. The end-to-end latency will be better by 10 folds by decreasing it to just 1ms. The battery lives of the 5G devices are going to be as long as 10 years a s both the energy efficiency and battery qualities are going to be better. Machine to machine communication will be much better than current 4G. Overall, the network and services reliability will be enhanced through advanced mechanisms such as software defined networking.

In order to provide further clarity on the 5G specifications, we show the ITU specifications in Table I. Right now, all the above mentioned performances along with the throughput and data rates proposed for 5G looks possible only through the optical communication systems. That makes optical communication an obvious and integral part of the 5G technologies. The recently tested 5G systems of Verizon and Samsung too use the optical networks for the core part and the access has to be wireless.

### III. Core-Access Hybrid Architecture

Before In this section, we present the hybrid nature of the proposed architectures. The main aim is to show the importance of optical communication in the 5G architecture. We propose an optical wireless hybrid system for 5G in which the major information carrying parts will be optical and the final hop between the base station and the end costumer will be wireless. This proposed architecture framework may be modified for different operating environments depending on the urban and rural setups.

For the urban environments we propose mm-wave based small cell architecture supported by high speed optical fiber connected base stations. In this case, for both the front-haul and back-haul up to the base stations (including the rest part of the core network) will be an optical network. In case of rural environment where the throughput and data rate demands are not very high as the urban will be served through high speed passive optical networks (PONs).

In Fig. 1, we show the architecture for urban environment. In this case, the core network is completely optical with all basic facilities for signal processing at the core nodes (also known as central gateways or CG). The core network is a ring which interconnects all the CGs with each other. Each core node is connected with several regional nodes (known as regional gateways or RG). This is how the regional rings are formed by interconnecting all the RGs with each other. The link capacities of the core networks are much higher than the link capacities of the regional networks. Then the local networks are formed as the tributaries of the regional

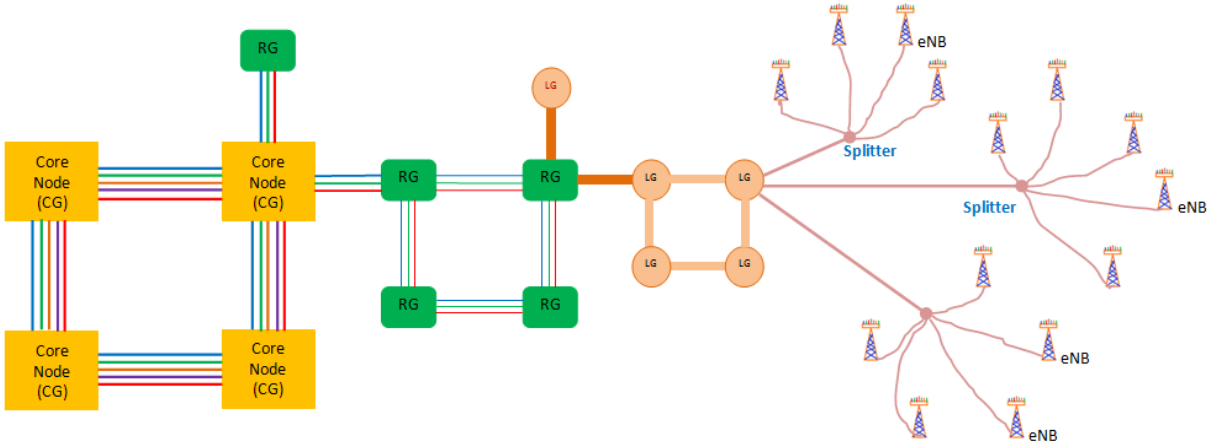
Fig. 2. 5G architecture for rural areas using a PON based system for the local areas. In this case, the whole architecture is optical except for the last mile.

networks. Each RG is connected with several local or metro network nodes (known as local gateways or LG). The LGs are interconnected with each other to form the local or metro rings. The LGs serve as the nearest optical processing hub for the bases stations or eNodeBs (eNBs). The connections between the LGs and the eNBs are also optical and the fiber connectivity is provided till the eNBs. All the eNBs are provided with the optical-to-electrical (OE) and electrical-to-optical (EO) facilities for the forward traffic and the backward traffic respectively. From the eNBs to the end users the communication is wireless.

So in this architecture, the core, regional and the local (or metro) networks are completely optical. For high speed and low latency the core and the regional networks can be made completely transparent (no need of any OE and EO conversions). Only the last mile is made wireless. In Fig. 2, we show the architecture for rural environment. In this case, the network architecture for the core, region and the local parts are very much same as the previous one shown in Figure 1. The only differences are found at the end of the LGs and the last mile scenarios. The eNBs are not directly connected with the LGs as in the previous case. Rather, they are connected through a PON. The splitter of the PON separates the incoming data stream and then sends them to different eNBs. This is required in the cases where the data rate demands are not very high. That is normally found in the rural areas. This PON based architecture is economical for the sparse traffic conditions. Therefore, it matches with the budget and requirements of the rural environment.

## IV. CONCLUSIONS

5G is going to be a complex heterogeneous network with several advanced features. Its architectures will also have several complexities. In the current situation, main features of 5G can be deliverable only with an active support of optical networks enabled with advanced technologies. The optical penetration in 5G much beyond what it used to be in 4G. The optical fibers in the urban environment are expected to be the feeders of the eNB. Only, in the very last hop it will be wireless and the length of the wireless hop will be much smaller than the previous generations. In our proposed architecture, we have shown that at least three layer of network hierarchy is needed for the 5G architecture. Two different types of architectures have been proposed for urban and rural environments. This is how the advanced features, complexities and the heterogeneity of the 5G networks can be handled.